\documentclass{article}
\usepackage{natbib,latexsym,epsfig,emulateapj,onecolfloat,graphics}
\def\eps@scaling{.95}
\def\epsscale#1{\gdef\eps@scaling{#1}}

\def\plotone#1{\centering \leavevmode
    \epsfxsize=\eps@scaling\columnwidth \epsfbox{#1}}

\def\kms{\ifmmode {\rm\,km\,s^{-1}}\else
    ${\rm\,km\,s^{-1}}$\fi}
\def\ms{\ifmmode {\rm\,m\,s^{-1}}\else
    ${\rm\,m\,s^{-1}}$\fi}
\def\kmsMpc{\ifmmode {\rm\,km\,s^{-1}\,Mpc^{-1}}\else
    ${\rm\,km\,s^{-1}\,Mpc^{-1}}$\fi}
\def\hkmsMpc{\ifmmode {\rm\,h^{-1}\,km\,s^{-1}\,Mpc^{-1}}\else
    ${\rm\,h^{-1}\,km\,s^{-1}\,Mpc^{-1}}$\fi}
\def\lya{{\rm Ly}$\alpha$}

\def\kpc{{\rm\,kpc}}

\def\msun{\ifmmode {\rm\,M_\odot}\else ${\rm\,M_\odot}$\fi}
\def\Msun{\ifmmode {\rm\,M_\odot}\else ${\rm\,M_\odot}$\fi}
\def\lsun{\ifmmode {\rm\,L_\odot}\else ${\rm\,L_\odot}$\fi}
\def\Lsun{\ifmmode {\rm\,L_\odot}\else ${\rm\,L_\odot}$\fi}
\def\rsun{\ifmmode {\rm\,R_\odot}\else ${\rm\,R_\odot}$\fi}
\def\Rsun{\ifmmode {\rm\,R_\odot}\else ${\rm\,R_\odot}$\fi}

\def\cmtw{\ifmmode {\rm\,cm^{-2}}\else ${\rm\,cm^{-2}}$\fi}
\def\cmthr{\ifmmode {\rm\,cm^{-3}}\else ${\rm\,cm^{-3}}$\fi}

\def\ergps{\ifmmode {\rm\,erg\,s^{-1}}\else ${\rm\,erg\,s^{-1}}$\fi}
\def\ergpscmtw{\ifmmode {\rm\,erg\,s^{-1}\,cm^{-2}}}

\def\eg{{\it e.g.}}

\def\deg{\ifmmode {^{\circ}}\else {$^\circ$}\fi}
\def\degr{\ifmmode {^{\circ}}\else {$^\circ$}\fi}
\def\degs{\ifmmode {^{\circ}}\else {$^\circ$}\fi}

\def\etal{{\it et al.~}}

\def\Ho{\ifmmode {\rm\,H_\circ}\else ${\rm\,H_\circ}$\fi}
\def\hnot{\ifmmode {\rm\,H_\circ}\else ${\rm\,H_\circ}$\fi}
\def\h0{\ifmmode {\rm\,H_\circ}\else ${\rm\,H_\circ}$\fi}
\def\hnotunit{\ifmmode {\rm\,km\,s^{-1}\,Mpc^{-1}}\else
    ${\rm\,km\,s^{-1}\,Mpc^{-1}}$\fi}
\def\qnot{\ifmmode {\rm\,q_\circ}\else ${\rm q_\circ}$\fi}
\def\q0{\ifmmode {\rm\,q_\circ}\else ${\rm q_\circ}$\fi}

\def\arcsec{\ifmmode {^{\prime\prime}}\else $^{\prime\prime}$\fi}
\def\asec{\ifmmode {^{\prime\prime}}\else $^{\prime\prime}$\fi}
\def\arcmin{\ifmmode {^{\prime}}\else $^{\prime}$\fi}
\def\amin{\ifmmode {^{\prime}}\else $^{\prime}$\fi}

\def\hone{H {\small I}}
\def\h{{\rm h}}
\def\lesssim{\mathrel{\hbox{\rlap{\hbox{\lower4pt\hbox{$\sim$}}}\hbox{$<$}}}}
\def\gtrsim{\mathrel{\hbox{\rlap{\hbox{\lower4pt\hbox{$\sim$}}}\hbox{$>$}}}}
\let\la=\lesssim                        
\let\ga=\gtrsim

\def\be{\begin{equation}}
\def\ee{\end{equation}}

\lefthead{C. V. Manning}
\righthead{Void Matter Density}

\begin{document}

\twocolumn [ 
\title{Birkhoff's Theorem and the Void Matter Density} 
\author{Curtis V. Manning}
\affil{Astronomy Department, University of California, Berkeley, CA
94720} 
\authoremail{cmanning@astro.berkeley.edu}

\begin{abstract}
According to the discussion on Birkhoff's theorem by
\citet{Peebles:93}, a void with a negative perturbation $\delta$ may
evolve as a separate homogeneous universe with a local expansion
parameter $H_V\simeq H_0(1-\delta/3)$.  This slightly low density
``universe'' will diverge from the mean, producing a void of ever
lower density.  As a result, the contents of voids will ``fall''
outward at a velocity equal to the difference between the local, and
the mean expansion parameters times the radius of the void.
Observational constraints on the outfall velocity can be placed as a
result of the fortuitous event that void, and non-void hydrogen \lya\
absorbers have distinct characteristics -- both in their equivalent
width distributions and their Doppler parameter distributions -- which
are clearly distinguished when cloud environments are measured in
terms of the sum of the tidal fields of surrounding galaxies acting on
the cloud \citep{Manning:02}.  These constraints dictate that the
radial outfall velocity from an average void of radius $\sim 15 \,{\rm
h}_{75}^{-1}$ Mpc must be less than $\sim100$ \kms.  The implications
of this are probed with a ``negative'' top-hat simulation, using a 1-D
Lagrangian code, to determine the relationship between the mass
deficit in voids and the local expansion parameter in a flat lambda
cosmology.  The outfall velocity constraint shows that the void matter
density must be greater than 75\% of the mean matter density.  I argue
that this implies that $\Omega_m \ga 0.86$.  Thus the total amount of
matter in voids is of order twice the total mass in the filamentary
structures.

\end{abstract}

]

\section{Introduction}

Voids are structures noted for their pronounced lack of galaxies;
concentrations of matter are thought to take the form of a luminous
``filamentary'' structure, so that together they resemble a ``foam''
of roughly spherical voids, edged by filaments and walls, and
sprinkled with richer concentrations of galaxies where the filaments
intersect.  While galaxies provide an important tracer of mass in the
universe, it is possible that there may be significant amounts of
undetected matter in the form of highly ionized clouds, or perhaps a
significant background of dark and gaseous matter, either within the
filamentary structure, or distributed within the voids.  Observations
of the low-redshift \lya\ cloud population have shown that clouds
cluster about galaxies \citep[\eg,][]{Tripp:98, Chen:01}, but there
are also many clouds which appear to be isolated.  

In \citet{Manning:02} (hereafter Paper 1), a primary cloud catalog
\citep{Penton:00b} was separated into complementary void and non-void
catalogs through the use of a parameter which assesses cloud
environments by weighting the effects of galaxy mass and distance.
This parameter, the scalar tidal field, which was summed over galaxies
within $7.5 ~\h_{75}$ Mpc of a cloud, is optimal for this purpose, as
it may easily be calculated from galaxy catalogs.  In addition, there
is a physical basis for supposing tides affect cloud distributions,
since it is the tide which determines the lower limit of cloud density
which is stable to dynamical disruption.  In Paper 1, a dimensionless
form of the tidal field ${\cal T}$ was evaluated at the locations of
each of a catalog of low-redshift \hone\ absorbers \citep{Penton:00b},
and used to characterize the clouds as members of a void, or non-void
environment, according to whether that field was greater or less than
some limiting tide ${\cal T}_{lim}$.  Equivalent width distribution
functions (EWDF) are calculated for each sub-catalog (see Eq. 44,
Paper 1).  These EWDFs are well-approximated as linear in $\log{d{\cal
N}/dz}$, \emph{vs.} $\log{\cal W}$, where ${\cal N}=n(\ge N_{HI})$,
and ${\cal W}$ is the rest equivalent width (EW) in m\AA.  One may
vary ${\cal T}_{lim}$, calculate the EWDFs of each sub-catalog, and
derive the weighted linear fits thereof, based on the number of clouds
contributing to the calculation of the EWDF at a given EW.
Ditsributions are fitted to the equation, $\log{d{\cal N}/dz}= {\cal
C} + {\cal S} \log{({\cal W}/63 \, {\rm m\AA})}$.  It was found that
the trends of fitting parameters for void and non-void catalogs as a
function of ${\cal T}_{lim}$ appear linear at large, or small ${\cal
T}_{lim}$ (see Fig. 1), but undergo a non-linear change within a
relatively small range of ${\cal T}$.  As it happens, there is a
corresponding change in the distributions of Doppler parameters at a
similar range of ${\cal T}$ for clouds sorted by ${\cal T}$ (see the
differential histogram, Fig. 8 in Paper 1).  It appears certain that
two distinct populations have been isolated, and that a transition
zone lies within the range, $-1.3 \la \log{\cal T} \la -0.7$.

\begin{figure}[h!] 
\centering
\epsscale{1.1} 
\vspace{-0.25in}
\plotone{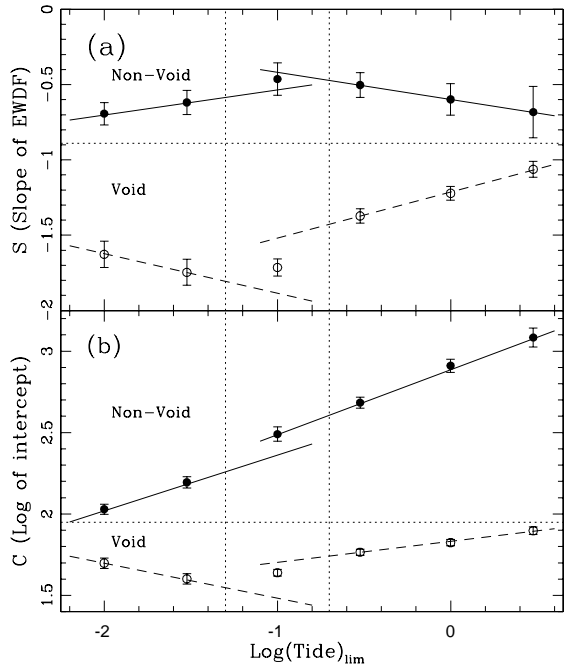}
\vspace{-0.1in}
\caption{The trend in slopes (panel $a$) and intercepts (panel $b$) of
equivalent width distribution functions defined by catalogs with tidal
field lower (top of each panel), or upper limits (bottom),
respectively, for the low-redshift cloud sample (see \S6.2.1 Paper 1).
There is an apparent strong transition in the slopes in the range
${-1.3 \la \log{\cal T}} \la -0.7$ (two dotted virtical lines), which
is also seen in the intercepts.}
\end{figure}

Apart from the interest in the detailed nature of these clouds, and
what they say about the nature of voids (paper in preparation), the
very fact that they can be sepated says something about the density of
voids.  For, by Birkhoff's theorem \citep{Birkhoff:23}, if voids have
a low density, they should behave like a low-density universe, and
suffer less deceleration, and have larger local expansion parameters,
than the mean.  With simulations I calculate the void expansion
parameter as a function of the void matter density relative to the
average.  If the smallness of the range of ${\cal T}$ over which the
transition of void to non-void clouds occurs can be used to constrain
the outfall velocity, then this can be used to constrain the the void
matter density.  This is, in short, my plan.

The cosmology assumed for this paper is a standard flat lambda model
with $h=0.75$.  The total matter density of voids is either referred
to as $\Omega_V={\bar \rho}_V/\rho_{crit}$, or as
$f_0=\Omega_V/\Omega_m$, where $\Omega_m = 0.3$ is initially assumed.
As noted in Paper 1, the large discovered line density of void clouds
appears to require that clouds are to a significant degree
self-gravitating and discrete.  I treat them as such herein.
\section{Constraints on Cloud Outfall Velocity}

Peculiar velocities in galaxy clusters cause a phenomenon known as the
``finger of God''.  A similar error in attributed position may result
when void clouds have large outfall velocities.  Large outfall
velocities will smear the distinction between void and non-void
clouds, as attributed positions are based on the redshift and the mean
Hubble constant (see \S1).  Figure 1 shows the trend of linear fitting
parameters for EWDFs defined by various ${\cal T}_{lim}$.  It appears
to show a fairly small amount of smearing, suggesting that the
transition zone (dotted vertical lines) is $-0.13 \la \log{\cal T}
\la -0.7$.  How can this observation be used to constrain the outfall
velocity?  The void filling factor as a function of ${\cal T}_{lim}$
provides the means.  Tidal fields were calculated along the pathlength
containing the absorption systems (see \S5 of Paper 1).  Figure 2
shows the fractional pathlength with ${\cal T} \le {\cal T}_{lim}$.
As a measure of the probability of a random spot in the universe
having a tide less than ${\cal T}_{lim}$, this is interpreted as the
fractional \emph{volume} with ${\cal T} \le {\cal T}_{lim}$.  Figures
4 in Paper 1 show that tidal fields increase rapidly at void edges.
By raising the tidal field upper limits, we are increasing the volume
of what we think of as voids.  The bracketing filling factors which
correspond to the above range of ${\cal T}_{lim}$ are, $0.75 \le f_V
\le 0.91$ (see Fig. 2), with a mean $\langle f_V \rangle=0.86$,
corresponding to ${\cal T}=0.1$.
\begin{figure}[h!] 
\centering
\epsscale{0.75} 
\vspace{-0.1in}
\plotone{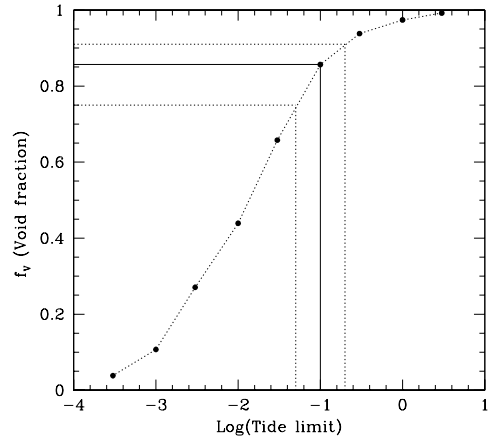}
\vspace{-0.1in}
\caption{The fraction of redshift space with ${\cal T} \le {\cal
T}_{lim}$ plotted against $\log{{\cal T}_{lim}}$.  This fraction can
be considered to be a fair approximation of the volume filling factor
for voids defined by ${\cal T} \le {\cal T}_{lim}$.  Dotted lines refer
to the region of transition similarly highlighted in Fig. 1.}
\end{figure}

Now, consider an average, fiducial void, which, at ${\cal
T}_{lim}=0.1$, has a radius of 15 Mpc ($\sim 11.25 \,{\rm h}^{-1}$
Mpc) \citep[e.g.,][]{Lindner:95}.  The range of uncertainty in the
position of the transition zone corresponds to radii of
$r_V=15\,(f_V/0.86)^{1/3} = 14.33$, and 15.396 Mpc, for void filling
factors $f_V=0.75$ and 0.91, respectively; an average variation of $
\pm 530 ~\kpc$.  This uncertainty in the true fiducial radius
corresponds to a line of sight (LOS) velocity error of $H_0 ~\delta r
\simeq \pm 40$ \kms, and would roughly reproduce the same range of
uncertainty apparent in the transition zone.  Admittedly, many clouds
will be moving along vectors at large angles to the LOS, and so even
with high outfall velocities, they could have low LOS peculiar
velocities.  But if clouds are randomly distributed in the void
relative to our LOS, and radially outfalling, an average of half of
the clouds would have to be moving at an angle less than 60 degrees
from the LOS, so that the observed LOS outfall velocity at the edge
would be greater than half of the \emph{radial} outfall velocity for
half of these clouds.  However, if clouds and mass are uniformly
distributed in voids, then half of the clouds will be at a radius $r
\ge 0.794 \, r_V$, so that most clouds would have an outfall velocity
$v_{out}\ga 80\%$ of the maximum.  Therefore, we expect that half of
the clouds will have a peculiar outfall velocity along the LOS $\ga
0.4$ times the radial outfall velocity at the void edge.  Since the
observed range of the LOS positional error corresponds to a velocity
error of $\sim 40$ \kms, this would seem to limit the true radial
outfall velocity at the void edge to be $v_{out} \la 40/0.4 = 100
\kms$.  However, since we know that some of the error in the
determination of the transition is plausibly due to peculiar
velocities of non-void clouds in proximity to galaxies (see \S5.2,
Paper 1), this is probably an over-estimation of the upper limit on
$v_{out}$.
\section{The Outfall Velocity from Voids}

Voids are thought to grow more rapidly than the scale factor
\citep[\eg,][]{Bertschinger:85a, Piran:97}.  This can be conceptually
understood by comparing the evolution of the expansion parameter for a
flat (the overall universe), and an open universe (the void).
  For the former,
\begin{equation}
H(z)=H_0\sqrt{\Omega_m(1+z)^3 + \Omega_{\Lambda}},
\end{equation}
and the latter, \citep{Scott:00},
\begin{equation}
H_V(z) = H_V(0) \sqrt{(f_0~\Omega_m
\,z+1-\Omega_{\Lambda})(1+z)^2+\Omega_{\Lambda}},
\end{equation}
where $H_V$ reflects a Hubble constant in the void, and $f_0$ is
defined in \S1.  One might think that as $z$ approaches infinity,
their respective expansion parameters would approach the same value so
that the ratio of their current values would simply be,
\begin{equation}
{\cal R}_H\equiv\frac{H_V(0)}{H_0}= \left(\frac{\Omega_m(1+z)^3 +
\Omega_{\Lambda}}{(f_0\,\Omega_m
\,z+1-\Omega_{\Lambda})(1+z)^2+\Omega_{\Lambda}} \right)^{1/2},
\end{equation}
so that a``peculiar'' outfall velocity could be determined by
subtracting off the mean Hubble flow,
\begin{equation}
v_{out}=({\cal R}_H-1)\,H_0\,r_V,
\end{equation}
where $r_V$ is the radius of the void.  The dashed line in Fig. 3
shows the functional form of Eq. 4 as a function of $f_0$.

However, while the theorem says that the lower-density void acts like
a separate ``universe'', our reconing of it, in terms of the
correspondence between lookback time and redshift, is not
straightforward, for the lookback time to a given redshift is a
function of cosmology, and a flat universe is younger than an open
universe with the same $H_0$: we cannot so easily derive our answer.

However, we may confidently approach this subject by simulating the
void as a ``negative'' top-hat with perturbation $\delta$.  A
1-dimensional Lagrangian hydro/gravity code \citep{Thoul:95} was
acquired courtesy of one of its authors, (A. A. T.), and adjusted for
a flat, $\Lambda$ cosmology (see \S1).  A more detailed description of
the code and some of its broader uses will be presented in a
subsequent paper (Manning, in preparation).  Linearized analysis of
voids dictate that the appropriate initial velocity perturbation
should be $v=(1-\delta/3)\,H_V(z) \,r$ \citep{Bertschinger:85a,
Peebles:93}, where $H_V$ is the local, void expansion parameter.
Simulations of the dark matter component were performed using this
velocity perturbation, with a range of uniform underdensities $\delta$
embedded in the otherwise flat universe, beginning at $z=50$.  A
$\delta=0$ case was run to normalize $H_0$.  The expansion parameters
at $z=0$ were derived.  The outfall velocity was calculated by
subtracting $H_0$ from $H_V$, and multiplying the result by the 15 Mpc
of our fiducial void.  Figure 3 shows the results (solid line) , where
$v_{out}$ is the outfall velocity.  For void densities $f_0 \la 0.2$,
outfall velocities are $v_{out} \ga 370 ~\kms$.  However, this
analysis (\S2) finds that $v_{out} \la 100 ~\kms$, and thus, $f_0 \ga
0.75$.  We may apparently conclude that the Hubble flow in voids is
very close to the average Hubble flow, and that therefore the
difference in density between voids and the mean is minimal.

\begin{figure}[h!] 
\centering
\epsscale{0.75} 
\plotone{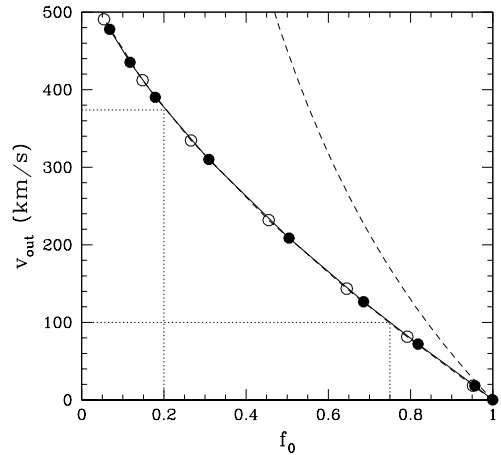}
\caption{The peculiar outfall velocity as a function of void density
relative to the mean for a void of radius $r_V=15$ Mpc, and ${\rm h} =
0.75$.  Results for a range of underdensities $\delta$ are shown as
filled circles for $\Omega_m=0.3$, and as open circles for
$\Omega_m=0.85$ (see \S4).  The dotted lines refer to characteristic
values of $f_0$ noted in the text.  The dashed line represents the
outfall velocity as derived from Eqs. 3 and 4, which ignores the
disparity between time-scales in flat and open cosmologies.}
\end{figure}

\section{Implications for $\Omega_m$}
Using the mean filling factors $f_V= 0.86$ for voids and $f_F =0.14$
for filaments, we may analyze the distribution of matter implied by
the above.
The basic equation for this analysis is,
\begin{equation}
\Omega_m = f_V \, f_0 \, \Omega_m + f_F\, \Omega_F,
\end{equation}
where $f_0 \, \Omega_m$ is substituted for $\Omega_V$.  Solving for
$\Omega_F$, when $\Omega_m=0.3$ and $f_0 \ge 0.75$, we find,
\begin{equation}
\Omega_F \la \frac{\Omega_m (1-f_V \, f_0)}{f_F} \simeq 0.761,
\end{equation}
which is absurd, since gravitational clustering could not occur unless
$\Omega_F > 1.0$.  It is apparent that $\Omega_F$ must be much
greater.  Introducing commonly accepted values into Eq. 6, for
instance, $\Omega_m=0.35$ and $f_0 = 0.18$, the mean of the range $0.1
\la f_0 \la 0.26$ \citep[\eg,][]{Cen:99,El-Ad:97b}, we find $\Omega_F
\simeq 2.2$.  If we assume this more reasonable value of $\Omega_F$,
then it turns out that using the above constraint on $f_0$, then
\begin{equation}
\Omega_m = \frac{f_F ~\Omega_F}{1-f_V \,f_0} \ga 0.86.
\end{equation}
That is, a low outfall velocity from voids mandates a ``high''
$\Omega_m$.  But is the outfall velocity for large $\Omega_m$ the same
function of $f_0$?  When the 1-D code was run for $\Omega_m=0.85$, a
graph consistent with that for $\Omega_m=0.3$ resulted (dashed line,
open points, Fig. 3), so the same constraints maintain in a high
$\Omega_m$ universe as with $\Omega_m=0.3$.
\section{Discussion and Summary}
\subsection{Impact on Filaments}
First, let us consider the effects of the deposition of energy in
volved in the outfall of void clouds onto filaments.  If clouds are
diffuse, and the kinetic energy of outfall is transformed entirely
into random motions, then $T=\mu m_H v_{out}^2/3 k$, where $k$ is the
Boltzmann constant, and $\mu \simeq 0.59$ is the mass of an average
particle.  A large velocity $v_{out}=400 ~\kms$ would produce a plasma
of temperature nearly $4 \times 10^6$ K, similar to that predicted by
\citet{Cen:99} for filaments, suggesting that absorbers should have
Doppler parameters $b \ga 250 ~\kms$, much larger than observed
$b$-values of \lya\ lines of non-void clouds ($b \approx 60 ~\kms$,
Paper 1).  For $v_{out}=100 ~\kms$, diffuse clouds are converted to a
temperature $T=2.4 \times 10^5$ K; $b\simeq 62 ~\kms$.  Interestingly,
\emph{Far Ultraviolet Spectroscopic Explorer} observations
\citep{Shull:00} find actual $b-$values may be roughly half that of
the observed \lya\ line, the excess being attributed to bulk motions.
Doppler parameters of order $30 ~\kms$ may be explained if clouds are
not diffuse, but held by dark matter halos, and hence centrally
condensed.  In this case, only gas with a density of order that of the
filament will be stripped, leaving the central cloud to proceed
relatively unmolested \citep{Murakami:94}, minimizing the thermal
broadening of the cloud.  Shear-induced vorticity \citep{Manning:99a},
will contribute to the Doppler broadening of the cloud, and ram
pressure will heat it, so that these absorbers should have greater
Doppler parameters than void clouds, as observed (Paper 1, \S5.6).
According to the scenario in \citet{Manning:99a}, the induced
vorticity of clouds, which are falling into dissipated gaseous
envelopes of galaxies and groups of galaxies, introduces an orderly
cycle of compression and cooling which results in elevated central
\hone\ equivalent widths as the cloud approaches mass concentrations.
This inverse correlation between cloud rest equivalent widths and
projected galactocentric radii has been observed \citep{Lanzetta:95,
Tripp:98, Chen:99,Chen:01}.

\subsection{Implications for Cosmology}
The conclusion that voids have a high matter density depends heavily
on the successful separation of void, from non-void clouds.  That the
Doppler parameter differential histograms, and trends in EWDF fitting
parameters both implicate the same transition zone suggests that this
is a robust result.  Outside the transition zone ${-1.3 \la \log{\cal
T}} \la -0.7$, the trends seen in Fig. 1 suggest distinct cloud
characteristics with few interlopers.  Within this range, however, the
trends change in a manner consistent with the effects of peculiar
velocities of non-void clouds about galaxies, compounded with that of
the outfall of clouds from voids with $v_{out} \la 100 ~\kms$ ($H_V
\la 1.1\ \, H_0$).  The simulations have shown that for $v_{out} \la
100 ~\kms$, the ratio of the void density to the mean is $f_0 \ga
0.75$, requiring $\Omega_m$ to be of order 1.  Under the strictures of
the flat Lambda model, this requires $\Omega_{\Lambda} \approx 0$, and
$\Lambda \approx 0$.

The current standard model postulates two forms of positive energy
which, under the analysis of the multipole fluctuations in the CMB,
appear to imply that $\Omega_m + \Omega_{\Lambda}\simeq 1$
\citep[\eg,][]{deBernardis:00}.  Independent assessments of $\Omega_m$
are all based on concentrations of matter, usually involving dynamical
tests, or biasing estimates \citep[\eg][]{Fukugita:98,Verde:01}, and
result in estimates $\Omega_m \approx 0.3$.  However, since these
methods are insensitive to a possible background, or additional
quantities of matter in void regions, these estimates are necessarily
\emph{under}-estimates.  As we have seen, constraints on the velocity
of outfall from voids imply a rather large ``background''.

On the other hand, supernova Ia studies strongly imply an exponential
expansion of the universe \citep[\eg,][]{Riess:98,Filippenko:01}.  How is
this to be reconciled?  What has not been shown by observation or
theoretical derivations is that what is created in conjunction with
the exponential expansion, and what enables the universe to be flat,
is a ``dark'' energy.  The conclusion that $\Omega_m\ga 0.85$ and
$\Lambda \gg 0$ can only mean that a dark energy is not formed, but in
its stead, ordinary matter, dark and baryonic.

It is unclear how one might adjust the parameters of the void
evolution simulations to be consistent with these conditions, for if
matter is forming in voids, the tendency of the mass density to
rapidly decline is reduced.  A careful numerical modeling of void
cloud absorbers as remnants of sub-galactic perturbations (paper in
preparation) may shed light on this problem (Manning, in preparation).

I thank Ann Thoul for the 1-D code, and Martin White, Mike Cai, and
Carl Heiles for useful discussions.  I am very grateful to Nahum Arav
for helpful advice on the manuscript.  I received financial support of
NSF grant \#AST-0097163 and the UC Department of Astronomy.




\bibliographystyle{apj} 

\end{document}


\bibitem[{{Scott}, D. and {Silk}, J. and {Kolb}, E..~W. and 
 {Turner},  M.~S.}]{Scott:00}
{Scott}, D. and {Silk}, J. and {Kolb}, E.~W. and {Turner}, M.~S. 2000, in 
 \emph{Allen's Astrophysical Quantities}, ed., {{Cox}, N.}, Springer-Verlag, 
  New York, Berlin, Heidelberg